\documentclass{article}
\usepackage{graphicx}
\usepackage{amssymb}

\begin{document}

\title{Transverse Spin at PHENIX: Results and Prospects}

\author{C.~Aidala for the PHENIX Collaboration \\
Columbia University \\
New York, NY 10027, USA \\
E-mail: caidala@bnl.gov}

\maketitle

\abstract{The Relativistic Heavy Ion Collider (RHIC), as the
world's first and only polarized proton collider, offers a unique
environment in which to study the spin structure of the proton. In
order to study the proton's transverse spin structure, the PHENIX
experiment at RHIC took data with transversely polarized beams in
2001--02 and 2005, and it has plans for further running with
transverse polarization in 2006 and beyond.  Results from early
running as well as prospective measurements for the future will be
discussed.}

\section{Introduction}

The Relativistic Heavy Ion Collider (RHIC) has opened up a new
energy regime in which to study the spin structure of the proton.
Polarization of more than 50\% has so far been achieved for 100-GeV
proton beams, with expectations that this value will rise to 70\% in
2006 or 2007.

The PHENIX experiment, one of two large experiments at RHIC,
specializes in the measurement of photons, electrons, and muons as
well as high-transverse-momentum ($p_T$) probes in general over a
limited acceptance, with good particle identification capabilities.
It has a high rate capability and sophisticated trigger systems,
allowing measurement of rare processes. The PHENIX
detector\cite{Adcox:2003zm} consists of two mid-rapidity ($|\eta | <
0.35$) spectrometers, primarily for identifying and tracking charged
particles as well as measuring electromagnetic probes, forward
spectrometers for identifying and tracking muons ($1.2 < |\eta| <
2.4$), and interaction detectors.

Several polarization-averaged cross sections have been measured for
200-GeV collisions at RHIC and found to be in good agreement with
next-to-leading-order (NLO) pQCD
calculations.\cite{Adler:2003pb,Adams:2003fx,Adler:2005in,Adler:2005qk}
The ability of NLO pQCD to describe RHIC cross section data well and
with little scale dependence provides a solid foundation for using
it to interpret polarized data in a similar kinematic regime.

\section{Current Results}

Large transverse single-spin asymmetries (SSAs) have been observed
in spin-dependent proton-proton scattering experiments spanning a
wide range of energies, as well as in semi-inclusive deep-inelastic
scattering. The origin of these asymmetries remains unclear, but
several different mechanisms have been proposed, as described for
example in Refs. \cite{Sivers:1989cc}--\cite{Kanazawa:2000hz}.

\begin{figure}[ht]
\includegraphics[width=0.95\linewidth]{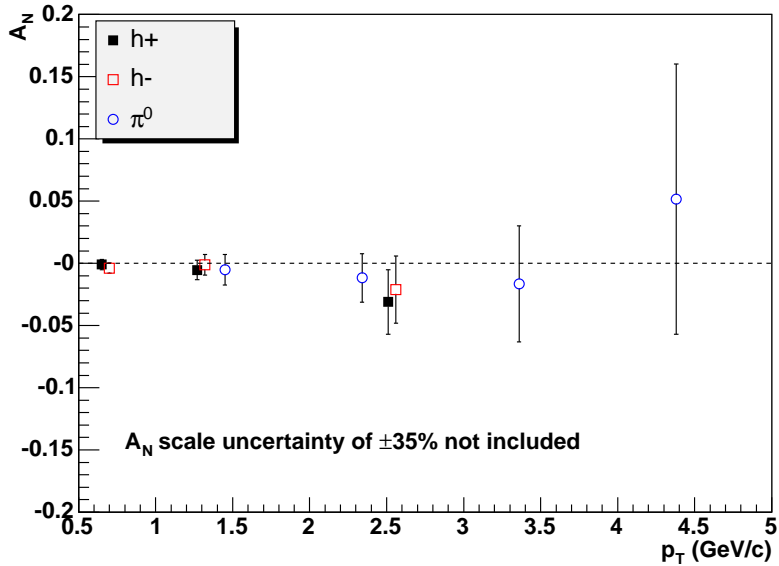}
\caption{ Mid-rapidity neutral pion and charged hadron transverse
single-spin asymmetries. Points for positive hadrons have been
shifted down by 50 MeV/$c$ to improve readability. The error bars
represent statistical uncertainties.}
\label{figure:finalAsymChargedNeutral}
\end{figure}

From data collected in 2001--02 (0.15~$\textrm{pb}^{-1}$, $\langle
P_{\textrm{beam}} \rangle \sim 15\%$), PHENIX measured the
left-right transverse single-spin asymmetry ($A_N$) for neutral pion
and charged hadron production at $x_F \sim 0.0$ up to a transverse
momentum of 5~GeV/$c$ from polarized proton-proton interactions at
$\sqrt{s} = 200$~GeV.\cite{Adler:2005in}  As can be seen in
Fig.~\ref{figure:finalAsymChargedNeutral}, the asymmetries observed
for production of both neutral pions and inclusive charged hadrons
are consistent with zero within a few percent over the measured
$p_T$ range.

\vspace*{-2mm}

\begin{figure}[ht]
\includegraphics[width=0.95\linewidth]{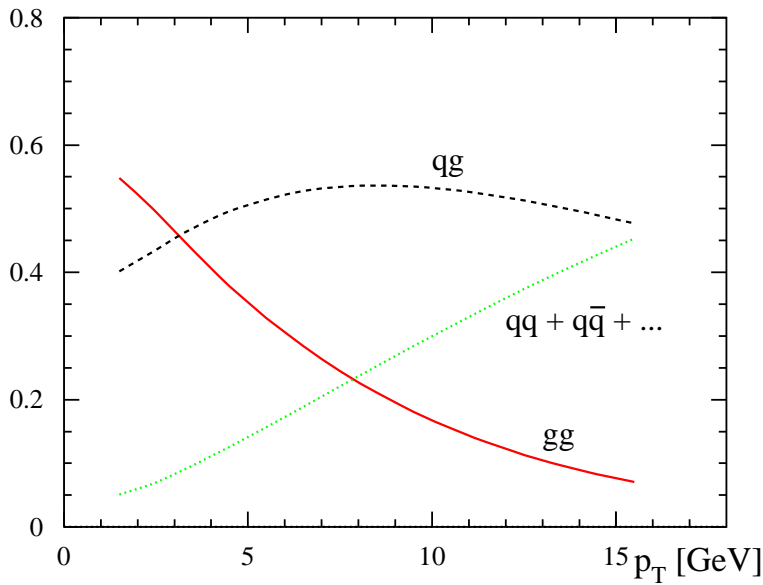}
\vspace*{-4mm} \caption{Relative fractional contributions of
partonic processes to mid-rapidity $\pi^0$ production at $\sqrt{s} =
200$~GeV, calculated by W. Vogelsang.}
\label{figure:partonicProcesses}
\end{figure}

The result is consistent with mid-rapidity results for neutral pions
at $\sqrt{s} = 19.4$~GeV.\cite{Adams:1994yu} The present measurement
is complementary to that of Ref.~\cite{Adams:2003fx}, which
observed large asymmetries for forward neutral pions at $\sqrt{s} =
200$~GeV. Neutral pion production at forward rapidity is expected to
originate from processes involving valence quarks, whereas particle
production at mid-rapidity is dominated by gluon-gluon and
quark-gluon processes. As evident from
Fig.~\ref{figure:partonicProcesses}, $\pi^0$ production in the $p_T$
range covered by the recent PHENIX measurement is nearly half from
gluon-gluon scattering and half from gluon-quark scattering. As
such, the asymmetry is not very sensitive to mechanisms involving
quarks.
\vspace*{-10mm}

In the forward direction at PHENIX, a large negative transverse SSA
of approximately -11\% in the production of neutrons from 200-GeV
$p+p$ collisions has been observed.\cite{Bazilevsky:2003} This
measurement was made using the RHIC zero-degree calorimeters (ZDCs),
hadronic calorimeters covering $2\pi$ in azimuth and $4.7 < |\eta| <
5.6$.  In 2005 there was a brief period of accelerator commissioning
with polarized proton collisions at 410~GeV, and the large negative
asymmetry in forward neutron production was found to persist.

The azimuthal asymmetry of forward charged particles was also
measured at PHENIX\cite{Taketani:2004} using beam-beam counters
(BBCs), which are quartz \v{C}erenkov counters that cover $2\pi$ in
azimuth and $3.0 < |\eta| < 3.9$.  The asymmetry for inclusive
forward charged particles was consistent with zero.  However,
non-zero asymmetries were found in charged particle production from
events in which a forward neutron was also detected in the ZDC. A
significant negative asymmetry was observed for forward charged
particles in neutron-tagged events, with a preliminary value of
$(-4.50 \pm 0.50 \pm 0.22) \times 10^{-2}$.  A smaller positive
asymmetry was found for backward charged particles produced in
neutron-tagged events, with a preliminary value of $(2.28 \pm 0.55
\pm 0.10) \times 10^{-2}$.  The observed asymmetries for forward and
backward charged particles in events with a forward neutron may
suggest a diffractive process.

\section{Prospective Future Measurements}
Despite great theoretical progress in recent years, no single, clear
formalism has emerged in which to interpret the currently available
data. Further theoretical work and a variety of additional
experimental measurements are necessary to understand current
results and elucidate the transverse spin structure of the proton.

From a modest transverse-spin data sample taken in 2005
(0.16~$\textrm{pb}^{-1}$, $\langle P_{\textrm{beam}} \rangle \sim
48\%$), PHENIX has begun analysis to obtain improved mid-rapidity
$A_N$ results for neutral pions and charged hadrons, expected to
provide tighter constraints on the gluon Sivers function. Future
higher-statistics samples for these particles at mid-rapidity will
reach higher $p_T$ and provide greater sensitivity to transversity
and the Collins effect.

There is also analysis underway to obtain first results for $A_N$ of
single muons, largely from open charm decay but with significant
contributions from light-hadron decays. The current $x_F$ reach for
this measurement is up to $\sim 0.15$; higher $x_F$ values would
become accessible with lower-energy running.  A forward hadron $A_N$
measurement using the PHENIX muon spectrometers may be possible
using decay muons and the charged hadrons that punch through the
absorber in front of the muon tracker. Careful studies will be
needed to understand the particle ratios in this sample.

In 2003 Boer and Vogelsang proposed a single transverse-spin di-jet
measurement that could probe the gluon Sivers
function.\cite{Boer:2003tx} A non-zero Sivers function implies a
spin-dependence in the $k_T$ distributions of the partons within the
proton, which would lead to an observable spin-dependent asymmetry
in $\Delta \varphi$ of back-to-back jets. In 2006, PHENIX intends to
perform a measurement similar to the one proposed, using di-hadrons
instead of di-jets because of the limited detector acceptance.  This
analysis will study the spin-dependence of the azimuthal angle
between nearly back-to-back $\pi^0$-hadron pairs, triggering on a
decay photon from the $\pi^0$ in order to obtain a higher-statistics
sample. Although dilution of the effect is anticipated for hadron
rather than jet pairs, studies have shown that it should still be
measurable. Fragmentation to the final-state hadrons must also be
considered, and some contribution from the Collins mechanism may be
present; however, as shown above in
Fig.~\ref{figure:partonicProcesses}, for $p_T \lesssim 5$~GeV/$c$
there is a large contribution to mid-rapidity $\pi^0$ production
from gluon fragmentation, to which the Collins mechanism does not
apply.

Measurement of $A_N$ for direct photons has also been proposed to
probe the gluon Sivers function.\cite{Schmidt:2005gv} Direct photon
production is dominated by quark-gluon Compton scattering ($q + g
\rightarrow \gamma + X$) over a wide range in photon $p_T$ at RHIC.
Transverse SSAs of photons and jets in events with correlated
photon-jet pairs would access the gluon and quark Sivers functions,
respectively, with some ability to identify the $x$ values at which
these functions were probed. PHENIX can currently measure $A_N$ of
mid-rapidity direct photons. Future upgrades extending the azimuthal
coverage for tracking to $2\pi$ in the inner region and adding
forward electromagnetic calorimetry ($0.9 < |\eta| < 3.0$) are
expected to expand the coverage for this measurement as well as make
$\gamma$-jet and jet-jet measurements feasible.

Yet another proposal has been made to access the gluon Sivers
function via mid- to moderate rapidity ($-0.2 < x_F < 0.6$) $D$
meson production at RHIC.\cite{Anselmino:2004nk} PHENIX is currently
capable of measuring open charm decays statistically via inclusive
single electrons and muons. In the future, a silicon vertex detector
upgrade will make it possible to identify $D$ mesons event by event.
Note that $A_N$ measurements for charmonium production, also
sensitive to the gluon, are already possible at PHENIX.  However,
the charmonium production mechanism is not as well understood.

The flavor separation of the Sivers function for $u$, $d$,
$\bar{u}$, and $\bar{d}$ quarks via $A_N$ of forward or backward $W$
boson production, possible once RHIC achieves 500-GeV collisions,
has been suggested by Schmidt.\cite{Schmidt:2005} The processes of
interest at PHENIX are $u + \bar{d} \rightarrow W^+ \rightarrow
\mu^+ + \nu_\mu$ and $d + \bar{u} \rightarrow W^- \rightarrow \mu^-
+ \bar{\nu}_\mu$. An upgrade to trigger on the high-$p_T$ muons from
$W$ decays is expected in 2009. The trigger upgrade will also make
open charm, charmonium, and Drell-Yan measurements cleaner.

The double transverse-spin asymmetry, $A_{TT}$, is another
observable sensitive to transverse spin quantities. $A_{TT}$ for the
Drell-Yan process would provide direct access to transversity.
Although this asymmetry is expected to be at the sub-percent level
for $\sqrt{s} = 200$~GeV, it could reach several percent for
$\sqrt{s} < 100$~GeV. PHENIX already has an effective di-muon
trigger for measuring Drell-Yan pairs; however, the trigger upgrade
will improve backgrounds. To measure $A_{TT}$ it would be necessary
to optimize the beam energy to balance luminosity against the size
of the predicted asymmetry.  A first direct measurement of
transversity would be an exciting milestone.

\section{Summary}
The first transverse-spin results from PHENIX are now available, and
further results from the brief transverse-spin run in 2005 are
forthcoming.  A longer period of running with transversely polarized
beams is anticipated for 2006.  Looking farther ahead, forward
detector upgrades will improve access to the kinematic region where
large asymmetries have been observed, and mid-rapidity upgrades will
improve jet measurements.

\section*{Acknowledgments}
PHENIX acknowledges support from the Department of Energy and NSF
(U.S.A.), MEXT and JSPS (Japan), CNPq and FAPESP (Brazil), NSFC
(China), MSMT (Czech Republic), IN2P3/CNRS, and CEA (France), BMBF,
DAAD, and AvH (Germany), OTKA (Hungary), DAE (India), ISF (Israel),
KRF, CHEP, and KOSEF (Korea), MES, RAS, and FAAE (Russia), VR and
KAW (Sweden), U.S. CRDF for the FSU, US-Hungarian NSF-OTKA-MTA, and
US-Israel BSF.

\end{document}